\documentclass {article}

\begin{document}

\def\oti{{\otimes}}
\def\bra#1{{\langle #1 |  }}
\def\lb{ \left[ }
\def\rb{ \right]  }
\def\tilde{\widetilde}
\def\bar{\overline}
\def\hat{\widehat}
\def\*{\star}

\def\({\left(}		\def\BL{\Bigr(}
\def\){\right)}		\def\BR{\Bigr)}
	\def\BBL{\lb}
	\def\BBR{\rb}
%
%

\def\bb{{\bar{b} }}
\def\ab{{\bar{a} }}
\def\zb{{\bar{z} }}
\def\zbar{{\bar{z} }}
\def\frac#1#2{{#1 \over #2}}
\def\inv#1{{1 \over #1}}
\def\half{{1 \over 2}}
\def\d{\partial}
\def\der#1{{\partial \over \partial #1}}
\def\dd#1#2{{\partial #1 \over \partial #2}}
\def\vev#1{\langle #1 \rangle}
\def\ket#1{ | #1 \rangle}
\def\rvac{\hbox{$\vert 0\rangle$}}
\def\lvac{\hbox{$\langle 0 \vert $}}
\def\2pi{\hbox{$2\pi i$}}
\def\e#1{{\rm e}^{^{\textstyle #1}}}
\def\grad#1{\,\nabla\!_{{#1}}\,}
\def\dsl{\raise.15ex\hbox{/}\kern-.57em\partial}
\def\Dsl{\,\raise.15ex\hbox{/}\mkern-.13.5mu D}
\def\b#1{\mathbf{#1}}
%
%
\def\th{\theta}		\def\Th{\Theta}
\def\ga{\gamma}		\def\Ga{\Gamma}
\def\be{\beta}
\def\al{\alpha}
\def\ep{\epsilon}
\def\vep{\varepsilon}
\def\la{\lambda}	\def\La{\Lambda}
\def\de{\delta}		\def\De{\Delta}
\def\om{\omega}		\def\Om{\Omega}
\def\sig{\sigma}	\def\Sig{\Sigma}
\def\vphi{\varphi}
%
%
\def\CA{{\cal A}}	\def\CB{{\cal B}}	\def\CC{{\cal C}}
\def\CD{{\cal D}}	\def\CE{{\cal E}}	\def\CF{{\cal F}}
\def\CG{{\cal G}}	\def\CH{{\cal H}}	\def\CI{{\cal J}}
\def\CJ{{\cal J}}	\def\CK{{\cal K}}	\def\CL{{\cal L}}

\def\CM{{\cal M}}	\def\CN{{\cal N}}	\def\CO{{\cal O}}
\def\CP{{\cal P}}	\def\CQ{{\cal Q}}	\def\CR{{\cal R}}
\def\CS{{\cal S}}	\def\CT{{\cal T}}	\def\CU{{\cal U}}
\def\CV{{\cal V}}	\def\CW{{\cal W}}	\def\CX{{\cal X}}
\def\CY{{\cal Y}}	\def\CZ{{\cal Z}}

\def\rvac{\hbox{$\vert 0\rangle$}}
\def\lvac{\hbox{$\langle 0 \vert $}}
\def\comm#1#2{ \BBL\ #1\ ,\ #2 \BBR }
\def\2pi{\hbox{$2\pi i$}}
\def\e#1{{\rm e}^{^{\textstyle #1}}}
\def\grad#1{\,\nabla\!_{{#1}}\,}
\def\dsl{\raise.15ex\hbox{/}\kern-.57em\partial}
\def\Dsl{\,\raise.15ex\hbox{/}\mkern-.13.5mu D}
\def\beq{\begin {equation}}
\def\eeq{\end {equation}}
\def\to{\rightarrow}

\title{Towards a unification of physics and information theory}

\author{I. Devetak\footnote{Electronic address: id21@cornell.edu} \\
\it{IBM T.J. Watson Research Center, Yorktown Heights, NY 10598} \\
\\
 A. E. Staples \\
\it{Department of Mechanical and Aerospace Engineering}\\ 
\it{Princeton University, Princeton, New Jersey 08540}
 }

  \date{\today} 
  \maketitle

\begin{abstract}
  
A common framework for quantum mechanics, thermodynamics and information theory is presented.
It is accomplished by reinterpreting  the mathematical formalism
of Everett's many-worlds theory of quantum mechanics  and augmenting it to include preparation
according to a given ensemble.
The notion of \emph{directed entanglement} is introduced through which
both classical and quantum communication over quantum channels
are viewed as entanglement transfer. This point is illustrated by
proving the Holevo bound and quantum data processing inequality relying
exclusively on the properties of directed entanglement.
Within the model, quantum thermodynamic entropy is also related to directed
entanglement, and a simple proof of the second law of thermodynamics is given.

\end{abstract}

\vspace{3mm}

We present a novel framework for quantum mechanics which we
argue to be more conducive to the development of quantum information theory.
Our motivation is both practical and one of principles.
On the practical side, we find the standard Copenhagen framework to be
unsatisfactory in various ways. For example, there are many different reasons for
a quantum system $Q$ to be
described by some mixed density operator $\rho_Q$.
It could be prepared
from some ensemble of pure states, it could be a result
of an unobserved measurement performed on a pure state, it
could be entangled with some other physical system, or it could be
some combination of these three. A related issue is that classical
and quantum communication over quantum channels appear to be rather
independent problems \cite{nie&chuang}. Another dichotomy is that
two protagonists may share information about a physical event
through \emph{communication} rather than through common observation,
yet these are treated very differently.

Regarding principles, we would like to view the laws of physics in information
theoretical terms. This attitude may be traced to Wheeler's motto "It from Bit" \cite{itbit},
and Wigner's observation that physics merely describes correlations between events \cite{Jahn}.
The two are unified  by the statement  that \emph{the totality of
conceptual experience can be described in terms of correlated random variables};
this will allow us to make contact with Shannon's information theory \cite{cover}
in which random variables are the carriers of information.
For instance, two protagonists sharing the same physical world is no 
more than classical correlations between the states of their knowledge regarding 
that world. Similarly, the observation of definite physical laws is no more than
classical correlations between states of knowledge regarding two consecutive acts 
of measurement, or preparation and measurement, depending on the experiment. 
A ball kicked by Alice seen as obeying Newton's deterministic laws of motion
is merely a statement about the correlation between her knowledge of its
initial velocity (preparation) and that of its position when it hits the ground
(measurement). 

The model we present here curtails the deficiencies of the Copenhagen
interpretation mentioned above, while embodying this information-theoretical principle.
Perhaps surprisingly, this can be accomplished by reinterpreting the mathematical 
formalism of a radically different theory of quantum mechanics, Everett's
Many-Worlds interpretation \cite{many}, and augmenting it to include 
\emph{preparation}, an indispensable ingredient for describing communication 
\cite{nie&chuang}.   
In our model the universe $\CU$ is divided into subsystems which can either belong
to the set of "physical" entities $\CP$ or to the set of instances 
of knowledge ${\cal K}$.  The set ${\cal K}$ partly corresponds 
to what Everett called
"observers", by which he essentially meant another physical system,
such as a quantum computer or the brain. Assuming that observers are 
entirely "physical" leads to the unsolved basis problem \cite{stapp}. 
Here we do not imbue ${\cal K}$ with a "physical" interpretation;
it merely keeps track of the conceptual experience of the 
observer/preparer in relation to what has been observed/prepared.
We pause to justify the unconventional step of
including conceptual experience in a physical model.
Everett points out in \cite{many} that physical 
theories have always consisted
of two parts, the "formal" mathematical part and the "interpretive" part: 
a set of rules for connecting the mathematical formalism to our conceptual
experience. He warns that the unrigorous nature of the second part often leads to 
a successful model being mistakenly identified with "reality." 
It is thus the conceptual experience that is the subject of physics,
and including it in the model itself can only help avoid the above pitfall. 

The set ${\cal K}$ is divided into subsets associated with particular
protagonists, such as ${\cal K}_{Alice}$ and ${\cal K}_{Bob}$.
At any given time the universe is described by a pure state $\ket{\Psi}_{\CU}$. 
The state of some subsystem $A$ is described by the density matrix obtained 
from $\ket{\Psi}_{\CU}$ by tracing out the rest of the universe ${\CU}/A$:

\beq
 \rho_A = tr_{\CU/A} \ket{\Psi} \bra{\Psi}_\CU.
\label{trace}
\eeq 
The Hilbert space of $\CU$ is constantly being augmented  by 
new instances of knowledge, initially in some fixed pure state 
(although they immediately get entangled with already existing
memebers of $\CU$; as pure states they serve no function).
Otherwise, $\ket{\Psi}_\CU$ can only evolve from one moment to the next according
to some unitary operator $U$

\begin{eqnarray}
 \ket{\Psi}_\CU  \longrightarrow  U \, \ket{\Psi}_\CU, 
\end{eqnarray}
possibly entangling the different subsystems.
These are the
building blocks of the theory. Now we illustrate how
measurement, preparation and communication are described in terms of it. 

\vspace{1mm}

(i) {\bf Measurement.} The measurement process \emph{\`a la} Everett 
is described as follows. Bob wishes to perform an elementary measurement
on some $m$-dimensional physical system $L$ in the orthonormal basis 
$\{ \ket{j}_L \}$. Denote by the \emph{reference system} $R$ that subsystem of 
the universe $\CU$ which is entangled 
with $L$, so that $RL$ is in a pure state 
$$
\ket{\Psi}_{RL} = \sum_{j = 1}^m \la_j \ket{\phi_j}_R \ket{j}_L . 
$$
The \emph{unobserved} measurement
consists of the measurement apparatus $M$, initially in some pure state,
becoming entangled with $RL$ via some unitary operator acting on $LM$
only. The state of $RLM$ becomes
$$
\ket{\Psi}_{RLM} = \sum_{j = 1}^m \la_j \ket{\phi_j}_R \ket{j}_L \ket{j}_M, 
$$
where $\{ \ket{j}_M \}$ is an orthonormal basis for $M$.
The \emph{observed} measurement consists of the production 
of an $m$-dimensional system $B \in \CK_{Bob}$ in some pure state,
followed by a unitary transformation acting on $MB$ only.
This results in
$$
\ket{\Psi}_{RLMB} = \sum_{j = 1}^m \la_j \ket{\phi_j}_R \ket{j}_L \ket{j}_M \ket{j}_B , 
$$
where $\ket{j}_B$ form an orthonormal basis for the Hilbert space $\CH_B$ of $B$.
These should be thought of as shorthand notation for the mutually exclusive 
states of Bob's knowledge with respect to the observation of $M$, 
$\ket{j}_B  \equiv \ket{{\sf observe} \,\, M \,\, {\sf in \,\, the \,\, state }\,\, \ket{j}_M }_B$.
The density matrix $\rho_B$ of $B$, viewed in the $\ket{j}_B$ basis, has diagonal
elements $|\la_j|^2$.
We define the associated random variable $\b B$ as
$$
  {Pr}(\b B = j) =  \bra{j} \rho_B \ket{j}_B = |\la_j|^2.
$$
The Shannon entropy of $\b B$ is defined as 
$H(\b B) = - \sum_j {Pr}(\b B = j) \log {Pr}(\b B = j)$.
The crucial point is the following. We started off by
modelling the mutually exclusive states of Bob's knowledge
by an orthonormal basis for $B$. If $\rho_B$ were diagonal 
in this basis, it would have the natural interpretation
of $B$ being in the state $\ket{j}_B$  with probability 
${Pr}(\b B = j)$, and thus isomorphic to the random variable
$\b B$. However, in general off-diagonal elements do exist; 
the theory then postulates that Bob is blind to this fact, since 
it is not in accord with his classical probabilistic vision, referred to henceforth
as "diagonal vision." This is, of course, just a manner of speech; 
all that is being said is that only the diagonal elements
have an intepretation.  

One might think that if Bob's experience is not based in "reality" there should be
a discrepancy between his experience and that of others. This is not the case.
If his friend Charlie takes a look at the readout of $M$, a new system
$C$ will be produced in such a way that the joint system $RLMBC$ is now in the state
$$
\ket{\Psi}_{RLMBC} = \sum_{j = 1}^m \la_j \ket{\phi_j}_R \ket{j}_L \ket{j}_M \ket{j}_B \ket{j}_C, 
$$
with the $\ket{j}_C$ defined analogously to $\ket{j}_B$.
To compare Bob's and Charlie's experiences we restrict attention to the diagonal elements of 
the \emph{joint} density matrix $\rho_{BC}$ associated with the joint random variable $\b{BC}$. 
It can be easily verified that $\b{B}$ and $\b{C}$ are perfectly 
correlated, namely ${Pr}(\b B = \b C) = 1$. In information theoretical terms, we have
$$
  I(\b B; \b C) = H(\b B) =  H(\b C),
$$
where $I(\b B; \b C) =  H(\b B) +  H(\b C) - H(\b {BC})$
is the mutual information between $\b B$ and $\b C$.  The same happens when Charlie 
does not observe the readout of $M$, but instead measures
$L$ in the same basis $\{ \ket{j}_L \}$ with his own apparatus $N$.
Bob  can also  perform a \emph{generalized} measurement
on some  physical system $Q$. This is done by entangling $Q$ with 
$L$ via some unitary operator acting on $QL$, and 
subsequently performing an elementary measurement on $L$, as described above.
Note that the only mathematical difference between our treatment of measurement 
and Everett's lies in our focus on density matrices in $\CK$ as the source of 
random variables. However, preparation, our next topic, is altogether new in this context.

\vspace{1mm}

(ii) {\bf Preparation.} 

Alice wishes to prepare some  $Q \in \CP$ according to an
\emph{ensemble} of pure states $\{ (p_i, \ket{\psi_i}_Q): i = 1, \dots n \}$,
$\sum_{i = 1}^n p_i = 1$, by which we mean that the state $\ket{\psi_i}_Q$ 
is prepared with probability $p_i$. This can be accomplished by 

1) measuring $Q$ and  bringing it into some fixed state $\ket{0}_Q$,
thereby disentangling it from its original reference system.
 
2) performing conditional unitary operations according to the ensemble
$\{ (p_i, U_i): i = 1, \dots n \}$, such that $\ket{\psi_i}_Q = U_i \ket{0}_Q$.
   
\vspace{1mm}

We omit the description of process 1); suffice it to say that 
the end result is the system $A_1 Q$ being in the state 
$$\ket{\Psi}_{A_1 Q} =    \ket{0}_{A_1} \ket{0}_Q, $$
where $A_1 \in \CK_{Alice}$ and $\ket{0}_{A_1}$ represents
Alice knowing that $Q$ is in the state $\ket{0}_Q$.
Now the outcome of process 2) is the creation of  $A_2 \in \CK_{Alice}$
so that the state of $ A_1 Q A_2$ is 
\beq
\ket{\Psi}_{A_1 Q A_2} =  \ket{0}_{A_1} 
\sum_{i = 1}^n \sqrt{p_i} \ket{\psi_i}_Q  \ket{i}_{A_2},
\label{first}
\eeq
where the basis state $\ket{i}_{A_2}$ stands for Alice knowing that she has performed
the unitary operation $U_i$. Finally  $A \in \CK_{Alice}$ is produced
giving rise to 
\beq
\ket{\Psi}_{A_1 Q A_2 A} =  \ket{0}_{A_1} 
\sum_{i = 1}^n \sqrt{p_i} \ket{\psi_i}_Q  \ket{i}_{A_2} \ket{i}_{A},
\label{second}
\eeq
where the basis state $\ket{i}_{A}$ represents
Alice knowing that $Q$ is in the state $\ket{\psi_i}_Q$ \emph{after}
the application of the conditional unitary transformation.
The transition from (\ref{first}) to  (\ref{second})
resembles a primitive quantum computation  \cite{nie&chuang}. 
 
Upon preparation, the system $Q$ is in the mixed state
$
\rho_Q = \sum_{i = 1}^n p_i  \ket{\psi_i} \bra{\psi_i}_Q
$.
The density matrix $\rho_A$ of $A$, viewed in the $\ket{i}_A$ basis, has diagonal
elements $p_i$. Just as  with Bob, 
these diagonal elements represent the probabilities of
Alice experiencing the corresponding basis states and we associated them with
the random variable $\b A$.

\vspace{1mm}
 
(iii) {\bf Communication.} Communication from Alice to Bob takes place 
by Alice encoding a message by  preparing a physical system $Q$ 
and Bob subsequently performing a generalized measurement on it. Thus 
the procedures of (i) and (ii) are combined, with the composite system $QA_2 A$ 
now playing the role of the reference system $R$. 
Consequently $A$ and $B$ become entangled via $A_2QLM$. The diagonal elements
of their joint density matrix $\rho_{AB}$ are now associated with the joint random
variable $\b{AB}$. 
 The mutual information between what Alice prepared and
what Bob received is simply $I(\b A; \b B)$. Thus it can be read off very simply
from the joint density matrix of Alice and Bob.

\vspace{3mm}

\paragraph{Directed entanglement and quantum channels.}
The theory hitherto presented 
suggests that all communication can be viewed
as entanglement transfer. Initially the sender $A$ is entangled with $A_2 Q$ only. 
Gradually the entanglement is passed on through $L$, $M$ and finally to $B$, 
the receiver. Intuition suggests that the final entanglement between $A$ and $B$
cannot exceed the initial entanglement between $A$ and $Q$. In addition, 
one would expect Alice's and Bob's diagonal vision to further reduce their 
"experienced" entanglement.
We now make these ideas concrete by introducing the notion of 
\emph{directed entanglement}.

Directed entanglement $E(X \rightarrow Y)$ from the system $X$ to the system $Y$
is defined as

\beq
  E(X \rightarrow Y) = S(Y) - S(XY),
  \label{de}
\eeq
where $S(Y)$ is short for the von Neumann entropy $S(\rho_Y)$ of the density matrix 
$\rho_Y$,
$S(\rho_Y) = -tr ( \rho_Y \, log \, \rho_Y)$. $S(XY)$ is defined analogously.
The connnection between $E$ and entanglement was observed by Schumacher and Nielsen
\cite{nie&sch} when it first appeared in the guise of coherent information $I_c$.
$E(X \rightarrow Y)$ may also be viewed as the negative of the conditional entropy
$S(X|Y) = S(XY) - S(Y)$, a quantity investigated in some detail, albeit without 
physical context, in \cite{nie&chuang}.

We list  some useful properties of $E(X \rightarrow Y)$:

\vspace{2mm}
(a) $- S(X) \leq E(X \rightarrow Y) \leq S(X)$

\vspace{2mm}
(b) $E(X \rightarrow Y) \leq E(X \rightarrow YZ)$ 

\vspace{2mm}
(c) $E(XY \rightarrow Z) = E(X \rightarrow Z) + E(Y \rightarrow XZ)$

\vspace{2mm}
(d) $E(X \rightarrow YZ) \geq E(X \rightarrow Y) + E(X \rightarrow Z)$

\vspace{2mm}
(e) $E(XY \rightarrow ZW) \geq E(X \rightarrow Z) + E(Y \rightarrow W)$

\vspace{2mm}
(f) $E(X \rightarrow Y) \geq E(X \rightarrow Y^c) \geq E(X^c \rightarrow Y^c)$ 

\vspace{2mm}
(g) $E(X \rightarrow Y)$ is invariant under local unitary transformations 
$U_X \otimes U_Y$

\vspace{2mm}
(h) $- S(X) \leq E(X \rightarrow Y^c) \leq 0$

\vspace{3mm}
 
We omit the proofs, many of which can be found in \cite{nie&chuang}.
In (b) equality holds when $Z$ is disentangled from $XY$, and we shall refer to
this fact as property (b$'$). Similarly, in (e) equality holds when $XZ$ is disentangled
from $YW$, and we label this by (e$'$).
In (f) and (h), $X^c$ refers to the \emph{classicized} system $X$, stripped of its off-diagonal 
elements in some preferred basis. Note that $S(X^c) = H(\b X)$. 
We now use these properties to prove
the two main theorems of classical
and quantum information processing: the \emph{Holevo bound} and the 
\emph{quantum data processing inequality}.

Consider sending classical information over some noisy channel
$\CE: \rho_Q \rightarrow \CE(\rho_Q)$. This differs
from (iii) in that the system $Q$ gets entangled with some unobserved environment $E$
(via some unitary transformation $U_{QE}$) between the preparation and  measurement phases. 
The total system involved in the process is thus $AA_2QELMB$.
Just after the interaction with $E$ the system $AQ$ is described by the
density matrix 
$$\rho_{AQ} = \sum_{i = 1}^n \sqrt{p_i} \ket{i} \bra{i}_A 
\CE(\ket{\psi_i} \bra{\psi_i}_Q),$$ and hence
$$E(A \to Q) =  \chi - H(\b A),$$
where the \emph{Holevo quantity} $\chi$ is given by
$$\chi = S(\CE(\rho_Q)) -  
\sum_{i = 1}^n p_i  S(\CE(\ket{\psi_i} \bra{\psi_i}_Q)).
$$
Denoting by primes quantities calculated after the 
interaction with $LMB$, we have  
the following string of equalities and inequalities

$$E(A \to Q) = E(A \to QLMB) = E'(A \to QLMB) $$
$$
\geq E'(A \to B) \geq  E'(A^c \to B^c) = H(\b B) - H(\b {AB}).
$$
\vspace{1mm}
The first four relations are due to (b$'$), (g), (b), and (f) respectively.
This gives rise to the Holevo bound

\beq
 I(\b A; \b B) \leq \chi.
\label{hb}
\eeq
Equality is asymptotically achieved by block coding in the limit 
of large blocklength \cite{holevo}.

One can also send \emph{quantum} information over a noisy channel.
Consider the physical system $Q$ being sent through two noisy channels 
$\CE_1$ and $\CE_2$ consecutively. Thus $Q$, initially entangled with the reference
system $R$ only, gets entangled first with $E_1$ and then
with $E_2$ via some $U_{A E_1}$
and $U_{A E_2}$, respectively. 
We denote by primes quantities calculated after the 
interaction with environment $E_1$ and by double primes those calculated
after the interaction with $E_2$. 
Then we have, by (b),
$$ E''(R \to Q E_1 E_2) \geq E''(R \to Q E_2) \geq E''(R \to Q) .$$
Noting  $ E''(R \to Q E_1 E_2) = E(R \to Q E_1 E_2) = E(R \to Q )$
and $ E''(R \to Q E_2) = E'(R \to Q E_2) = E'(R \to Q)$, both
consequences of (g) and (b$'$), we get the quantum
data processing inequality

\beq
 S(\rho_Q) \geq I_c(\rho_Q, \CE_1) \geq I_c(\rho_Q, \CE_2 \circ \CE_1), 
\label{pipe}
\eeq
where $I_c$ is the coherent information \cite{nie&sch}.
Equality is again achieved asymptotically \cite{noisy}.

We have seen that the key quantum 
noisy channel relations for sending classical (\ref{hb}) and quantum (\ref{pipe})
information both follow from the properties of directed entanglement. 
\footnote{It should be stressed that the proofs given are mathematically akin to 
already existing ones \cite{nie&chuang}, only reexpressed in the common
language of entanglement transfer.}
We expect this to open avenues, e.g., for adapting the quantum channel coding techniques 
developed by Lloyd \cite{noisy} to produce classical codes saturating the Holevo bound,
just like Schumacher compression may be applied to classically prepared information.

\vspace{2mm}

\paragraph{Quantum thermodynamics.}
With the developments of the previous sections it
becomes clear that the \emph{thermodynamic entropy
can only be defined relative to elements of $\CK$}.
We define the thermodynamic entropy $S_T(Q|B)$ of a system $Q$ relative to 
$B \in {\cal K}_{Bob}$ as 
\beq
S_T(Q|B) =  S(Q|B^c) = - E(Q \to B^c),
\label{st}
\eeq
i.e. it is the negative of the directed entanglement from $Q$ to the classicized system $B$.
In terms of standard quantum mechanics, it corresponds to the average von Neumann
entropy of the system $Q$ "as viewed by Bob" upon measuring or preparing it.
By property (h), it lies between $0$ and $S(Q)$, the upper bound attained
when $B$ is disentangled from $Q$.

Of course, our definitions of the thermodynamic entropy must be justified by the
zeroth and second laws of thermodynamics \cite{partovi}.
For proving the zeroth law it
suffices that the sum of thermodynamic entropies of two initially unentangled systems
cannot decrease when they interact (see \cite{partovi} for a full treatment). 
This, in turn, follows from
properties (b) and (c), which give 
$ S_T(Q_1|B) + S_T(Q_2|B) \geq S_T(Q_1 Q_2|B)$, 
and noting that equality holds when $Q_1$ and $Q_2$ are unentangled.  

The second law is somewhat more subtle, and holds only for macroscopic systems. 
The Hilbert space of our macroscopic system  $Q$ may be divided into a tensor product
$\CH_Q = \CH_{Q^>} \otimes \CH_{Q^<}$, where $Q^>$ and $Q^<$ represent the macroscopic 
and microscopic degrees of freedom, respectively. The second law is a consequence of the
lack of knowledge of the microstructure of the physical system, and the
macrostructure being converted into microstructure as the system evolves
\cite{waldram}. Accordingly, it is the \emph{coarse-grained} entropy, 
$S^>_T(Q|B) \equiv S_T(Q^>|B)$, i.e. the thermodynamic entropy of $Q^>$,
that increases. To prove this we introduce a dummy system $\tilde{B} \in \CK_{Bob}$ 
which keeps track of Bob's (nonexistent) knowledge of $Q^>$, and hence remains disentangled
from $Q B$ at all times. Initially $B$ is maximally entangled with $Q^>$,
as a consequence of preparation or measurement,
and $Q^<$ has a uniform density matrix and is disentangled from $B Q^>$.
By property (e$'$) we have $E(Q^> Q^< \rightarrow B^c \tilde{B}) 
= E(Q^> \rightarrow B^c) + E(Q^< \rightarrow \tilde{B})$.
Since $Q$ is closed, it evolves via some unitary operator $U_{Q^> Q^<}$, 
entangling $Q^>$ and $Q^<$, so that $E'(Q^> Q^< \rightarrow B^c \tilde{B}) =
E(Q^> Q^< \rightarrow B^c \tilde{B})$ by (g). Property (e) implies 
$E'(Q^> Q^< \rightarrow B^c \tilde{B}) \geq E'(Q^> \rightarrow B^c) + 
E'(Q^< \rightarrow \tilde{B})$, and hence (in obvious notation)
$\Delta E(Q^> \rightarrow B^c) + \Delta E(Q^< \rightarrow \tilde{B}) \leq 0$.
Since $Q^<$ initially had a uniform density
matrix, it follows that $\Delta E(Q^< \rightarrow \tilde{B}) \geq 0$.
Therefore $\Delta E(Q^> \rightarrow B^c) \leq 0$,
i.e. $S^>_T(Q|B)$ has increased. There is, however, no a priori reason why $S^>_T(Q|B)$
should \emph{continue} to increase since the density matrix of $Q^<$ is not 
uniform any more. At this point we invoke the idea, central to Wilson's renormalization group,
that the many lengthscales of $Q$ are \emph{locally coupled} \cite{wilson}.
Hence we may further decompose the Hilbert space of $Q^<$ into spaces of increasing size,
corresponding to decreasing lengthscales, $\CH_Q =  \CH_{Q_0} \otimes \CH_{Q_1} \otimes \cdots $,
in such a way that the time evolution entangles only neighboring spaces, causing a cascade effect.
In the light of this principle, the above calculation is now modified by replacing 
$Q^<$ with $Q_0$. After $Q^>$ and $Q_0$ interact, $Q_0$ will interact with 
the completely uniform (and larger in size) $Q_1$, which will thoroughly randomize it.
Thus, when $Q^>$ and $Q_0$ interact again, $Q^<$ starts off with a uniform 
density matrix so that $S^>_T(Q|B)$ continues to increase. 
The cascade continues, the vast microscopic
degrees of freedom constantly radomizing the more macroscopic ones, until thermodynamic
equilibrium is reached. Interestingly, the same method works if we consider the second
law to be a consequence of interactions with an unobserved environment, rather than
coarse graining \cite{partovi2}. The microscopic degrees of freedom are, in essence,
an unobserved environment.

\vspace{2mm}

The model presented provides a simplified and unified view of physics and 
communication, with directed entanglement (\ref{de}) as the key binding 
quantity. As advertised, correlated random variables come about rather naturally,
and drawbacks of the Copenhagen interpretation mentioned in the introduction are 
taken care of. Mixed density operators
are \emph{always} a consequence of entanglement, and both classical and quantum 
communication over quantum channels are seen as the transfer of entanglement.
Correlated random variables describing 
the conceptual experience of various protagonists arise from the entanglement 
between different instances of knowledge, interpreted through 
diagonal vision. These instances of knowledge are
always entangled via some physical system. In (i) Bob and Charlie experienced the 
same measurement result due to being entangled via $Q$. Alternatively,  Charlie 
could have learned the measurement outcome by Bob communicating it to him, in which 
case they would get entangled indirectly via some physical system used to encode the message.
Finally, quantum thermodynamic entropy is defined as the lack of entanglement between
instances of knowledge and physical systems. In addition we have presented a simple
and intuitive proof of the second law of thermodynamics.
Thus the model promises to become a very practical framework for understanding and 
developing quantum information theory and quantum thermodynamics. 
On a  philosophical level it makes Bohr's insight 
that the wavefunction merely represents knowledge
much more precise by relating it to the \emph{true} 
carriers of Shannon information -- random variables.

\vspace{1mm}

{\bf Acknowledgments.} We are indebted to R. Maimon for constructive criticism
of earlier versions of the manuscript. We also thank 
Z. Had{\v{z}}ibabi\'c, J. Karcz, D. Milanov, M. Milosavljevi\'c 
and K. Narayan for useful discussions. I.D. acknowledges the partial support of 
the DoD Multidisciplinary University Research
Initiative (MURI) program administered by the Army Research Office under
Grant DAAD19-99-1-0215 while at Cornell University.


\begin{thebibliography}{99}

\bibitem{nie&chuang} M.A. Nielsen and I.L. Chuang,
{\it Quantum Computation and Quantum Information}, Cambridge U.P., Cambridge, UK (2001)




\bibitem{itbit} J. A. Wheeler  in {\it Complexity, Entropy and the Physics
of Information}, W. H. Zurek, ed., Addison-Wesley, Redwood City, CA (1991) 

\bibitem{Jahn} E. Wigner in { \it The Role of consciousness in the physical world},
 R. G. Jahn ed., Westview Press, Boulder, CO (1981)

\bibitem{cover} C. E. Shannon and W. Weaver,
{\it The mathematical theory of communication}, U. Illinois Press, Urbana, IL (1949)

\bibitem{many} H. Everett, III in {\it The Many-Worlds Interpretation
of Quantum Mechanics}, B.S. DeWitt, N. Graham, eds., Princeton U.P., Princeton, NJ (1973)

\bibitem{stapp} H.P. Stapp, quant-ph/0110148 (2001)
\bibitem{nie&sch} B. Schumacher and M. A. Nielsen {\it Phys. Rev. A} {\bf 54}, 2629 (1996)
  
\bibitem{holevo} A.S. Holevo, { \it IEEE Trans. Inf. Theory} {\bf 44(1)}, 269 (1996); 
B. Schumacher and M.D. Westmoreland,  {\it Phys.Rev.A }{\bf 56}, 131 (1998); 

\bibitem{noisy} S. Lloyd, 
 {\it Phys. Rev. A }{\bf 55},1613 (1996)

\bibitem{partovi} M.H. Partovi, {\it Phys. Lett. A }{\bf 137},
440, (1989)

\bibitem{waldram} J.R. Waldram,
{\it The Theory of Thermodynamics}, Cambridge U.P., Cambridge, UK (1985);
A. Wehrl, {\it Rev. Mod. Phys } {\bf 50}, 221 (1978)

\bibitem{wilson} K.G. Wilson, {\it Rev. Mod. Phys } {\bf 47},
773 (1975) 

\bibitem{partovi2} M.H. Partovi, {\it Phys. Lett. A }{\bf 137},
445, (1989)




\end{thebibliography}
\end{document}